%% file: main.tex
\documentclass[sigconf]{acmart}
\usepackage{subfig}
\usepackage{graphicx}
\setcopyright{none}
\renewcommand\footnotetextcopyrightpermission[1]{} 

\usepackage{color}

\AtBeginDocument{%
  \providecommand\BibTeX{{%
    \normalfont B\kern-0.5em{\scshape i\kern-0.25em b}\kern-0.8em\TeX}}}

\usepackage{pifont}
\usepackage{breqn}

\begin{document}

\title{Where Do You Want To Invest? Predicting Startup Funding From Freely, Publicly Available Web Information}


%
\author{Mariia Garkavenko}
\affiliation{%
 \institution{University of Grenoble Alpes}
 \institution{Skopai}
 \city{Grenoble}
 \country{France}}
\email{mariya.garkavenko@skopai.com}

\author{Eric Gaussier}
\affiliation{%
	\institution{University of Grenoble Alpes}
	\city{Grenoble}
	\country{France}}
\email{eric.gaussier@imag.fr}

\author{Hamid Mirisaee}
\affiliation{%
	\institution{Skopai}
	\city{Grenoble}
	\country{France}}
\email{hamid.mirisaee@skopai.com}

\author{C\'edric Lagnier}
\affiliation{%
	\institution{Skopai}
	\city{Grenoble}
	\country{France}}
\email{cedric.lagnier@skopai.com}

\author{Agn\`es Guerraz}
\affiliation{%
	\institution{Skopai}
	\city{Grenoble}
	\country{France}}
\email{agnes.guerraz@skopai.com}
%
%
%
%
%


\begin{abstract}
We consider in this paper the problem of predicting the ability of a startup to attract investments using freely, publicly available data. Information about startups on the web usually comes either as unstructured data from news, social networks and websites, or as structured data from commercial databases, such as Crunchbase. The possibility of predicting the success of a startup from structured databases has been studied in the literature and it has been shown that initial public offerings (IPOs), mergers and acquisitions (M\&A) as well as funding events can be predicted with various machine learning techniques. In such studies, heterogeneous information from the web and social networks is usually used as a complement to the information coming from databases. However building and maintaining such databases demands tremendous human effort.
We thus study here whether one can solely rely on readily available sources of information, such as the website of a startup, its social media activity as well as its presence on the web, to predict its funding events. As illustrated in our experiments, the method we propose yields results comparable to the ones making also use of structured data available in private databases.
\end{abstract}



\keywords{predictive modeling, heterogeneous web data, mining open sources, startup fundraising}

\maketitle
\pagestyle{plain}

\input{1Introduction}
\input{2Related}
\input{3Data}

\input{4Prediction}
\input{5Evaluation}

\input{6Conclusion}


\bibliographystyle{ACM-Reference-Format}
\bibliography{main}
\input{Appendix}

\end{document}

%% file: 1Introduction.tex
\section{Introduction}\label{sec:intro}
In recent years, startups have radically changed the situation in many different economical ecosystems and have become the pioneers of world-class innovations.
Furthermore, successful startups, significantly impact their targeted market and return a considerable profit to their investors.
Accordingly, Venture Capital firms (VCs) spend a considerable amount of time to identify and monitor startups.
However, many startups fail in the first few years \cite{hyytinen2015does} for different reasons: poor product-market fit \cite{feinleib2011startups}, improper managerial strategies \cite{giardino2014early} and more importantly inability to raise enough funds \cite{cremades2016art}. Receiving funds is indeed considered as one of the metrics for assessing the success of a startup and is a key element in the decision of future VCs and investors. However, the fact that different funding rounds usually appear within long time intervals makes the prediction task more challenging. In addition, unlike big companies, available information on the web on startups is rather limited since, as a young company, there is not much history about their products, services, etc.

With the recent advances in the machine learning, many problems have been reformulated as prediction tasks which can then be solved by ML approaches. The problem of predicting the success of a startup by predicting its future funding events has been investigated by the research community using different ML approaches (see \cite{zhang2017predicting, xiang2012supervised} for example). These studies have exploited standard features, as the size of a startup and its location, as well as social media-based features capturing the activity of a startup on social media. In particular, features extracted from Twitter have been shown to be crucial for predictive models, especially for predicting the ability to raise funds \cite{tumasjan2019gotta, antretter2019predicting}. In addition to these features, state-of-the-art methods for predicting funding events \cite{sharchilev2018web} rely on manually curated features available in proprietary databases such as Crunchbase\footnote{https://www.crunchbase.com/} and describing, \textit{e.g.}, the previous funding events of a given startup, with the amount raised, their type (as seed, angel or venture), the number of rounds without valuations, etc. Maintaining such a database is nevertheless a costly, time consuming task that is furthermore likely to be incomplete in the sense that it is very difficult to be exhaustive in the startups covered. For these reasons, we investigate here the possibility to predict funding events of startups by only resorting to features that can automatically be extracted from freely, publicly available sources of information.

More precisely we want to predict whether a startup will secure funding round in a given amount of time (\textit{horizon}) given its feature vector, which corresponds to a traditional binary classification task. 
Our contribution on this is twofold:
\begin{enumerate}
\item First, we show how one can extract a rich set of features describing startups from freely, publicly available sources of information as startup websites, social media and company registries;
\item Second, we show that, by using state-of-the-art ML methods with these features, one can obtain prediction results that rival the ones obtained with manually curated features.
\end{enumerate}

The remainder of this paper is organized as follows: Section~\ref{sec:related} provides an overview of  the most related studies. Section~\ref{sec:framework} describes the data and the feature extraction process. Section~\ref{sec:models} presents the ML frameworks we have investigated. Then in Section~\ref{sec:eval}, in order to illustrate the effectiveness of the proposed approach, we present a set of experiments. We also report and analyze, in the same section, the results of experiments. Finally, Section~\ref{sec:conclu} concludes the paper. 

%% file: 2Related.tex
\section{Related Work}\label{sec:related}
As explained in Section \ref{sec:intro}, the pace at which startups are created and the impact that they have on economy is both new and important, and relatively few studies have investigated the problem of predicting the trajectories of startups. 
The work presented in \cite{xiang2012supervised} is perhaps one of the first attempts to dive into the field of using predictive models for assessing the "success" of companies. In that work, the authors explore the prediction of Merger \& Acquisition (M\&A) as a metric to examine if a company should be categorized as successful or not. To do that, they consider news pertaining to companies and individuals on TechCrunch. The input space defined in this paper includes company-specific features, such as managerial  and financial features, combined with topic-dependent features which have been extracted via Latent Dirichlet Allocation (LDA).

In a relatively similar work, Hunter et al. \cite{hunter2017picking} proposed to construct a portfolio of startups in which at least one startup achieves an exit, \emph{i.e.} either gets Initial Public Offering (IPO) or is acquired by another company. The data studied in this work is principally taken from three sources: CrunchBase\footnote{\url{https://www.crunchbase.com/}}, Pitchbook\footnote{\url{https://pitchbook.com/}} and LinkedIn. Starting from a Brownian motion model, the authors propose to use a greedy approach to solve the "picking winners" problem. Our work differs form these two studies in that, firstly, the objective that we follow is basically prediction of funding events. Secondly, and more importantly, the data we use to train our models is publicly available, unlike LinkedIn data or VC databases, and comes from verified sources, \emph{i.e.} the startups themselves, while datasets like TechCrunch can be edited by anyone.

Over the past decade, social media, in particular Twitter, have been extensively used in building ML models. Event detection, sentiment analysis and success prediction are only a few examples of Twitter-based models \cite{chen2014topic, atefeh2015survey, li2016project}. Likewise, in our context, as mentioned in \cite{antretter2019predicting}, Twitter can play a crucial role in survival or failure of new ventures. Given the fact that tweets are publicly available and, if written by the company itself, are usually reliable sources of information, they can be used to predict the evolution of startups.

One such study that investigates the role of social media on the success of companies is \cite{zhang2017predicting}. Therein, it is shown that the social engagement of a startup, like the number of tweets and the number of followers, has a significant correlation with its success in receiving crowdfunding.
The authors further argue that, since crowdfunding is rather a new fundraising technique in which the traditional face-time approach with investors does not practically exist, the activity of a startup on social media such as Facebook and Twitter heavily boosts the odds of receiving crowdfunds. In that paper, AnglelList\footnote{\url{https://angel.co/}} (a crowdfunding investment platform), Facebook and Twitter were chosen as data sources, where AnglelList has been used to identify the startups which received crowdfunds. To select startups, the authors picked 4001 startups, on a snapshot of 744K, which were actively fundraising, where, after data cleaning, only 271 startups, with 11 positives, were kept to perform prediction.
The data being highly imbalanced, a wide range of techniques, such as \cite{castro2013novel, krawczyk2016learning, wu2017multiset}, combined with a greedy feature selection algorithm were used  in order to train models with better true positive rates (TPR). Through an extensive analysis, it is shown that the method presented in \cite{zhang2017predicting} can reach an accuracy of 84\% in predicting crowdfunding events. Compared to \cite{zhang2017predicting}, the present study not only exploits social media data, but also uses a series of other features to perform the fundraising prediction. As it will be shown in Section~\ref{sec:eval}, these features increase the prediction performance in a significant manner.

In a more recent study and following the same line, Sharchilev et al. \cite{sharchilev2018web} proposed a method, named Web-Based Startup Success Prediction (WBSSP),  for startup success prediction. This paper is perhaps the most relevant study to ours in the literature. The goal of \cite{sharchilev2018web} is to predict, within a period of time in the future, if a startup which already secured seed funding will get further rounds of funding or not. Their source of data, as the studies mentioned above, is CrunchBase, LinkedIn and the web. To construct the model, the authors define a feature space which is further decomposed into four categories: general, investor, team and mentions on the web. They showed that the last category, \emph{i.e.} web-based mentions, boost the prediction performance in their proposed pipeline. In order to avoid overfitting, the learning algorithm is designed by grouping the features into sparse and dense features. The features are then passed through a set of learning methods which are then fed into a boosting method (CatBoost \cite{prokhorenkova2018catboost}) in order to obtain the final prediction results. Eventually, the authors discussed a set of research questions, particularly the importance of each category of features and the effect of treating the sparse features differently than the dense ones.


The present work differs from \cite{sharchilev2018web} in two important aspects: first, we rely on features that can be extracted from freely, publicly available information that does not require costly, manual development and maintenance; second, we rely on simple machine learning models that can be easily implemented and reproduced. Overall, the features we rely on are simpler to extract, the prediction models we use easier to develop and the results we obtain at least as good as the ones presented in \cite{sharchilev2018web}.

%% file: 3Data.tex
\section{Data Collection}\label{sec:framework}
In this section, we describe in detail the process of collecting the data required to solve the prediction task. 
The first step in this process is to build a list of the startups for analysis, along with the links to their websites.
To do that, one can use any sufficiently large list of startups available on the web. In our case, we gathered 22K startups from multiple sources, such as hubs, investors and conferences, across the world.
Once having collected the list of the startups websites, we extract information from the following sources:
\begin{itemize}
	\item Startup's own website,
	\item Twitter API\footnote{\url{https://developer.twitter.com}},
	\item Google search API\footnote{\url{https://developers.google.com/}},
	\item Country-specific registration data on companies (\emph{e.g.} Infogreffe \footnote{\url{https://www.infogreffe.com}}) containing information about firms, such as the office locations and number of employees.\footnote{\label{after_acceptance}The complete list can be found in the Appendix~\ref{app:biz_reg}.}
\end{itemize}

\begin{figure}[t]
	\footnotesize
	\includegraphics[width=\columnwidth]{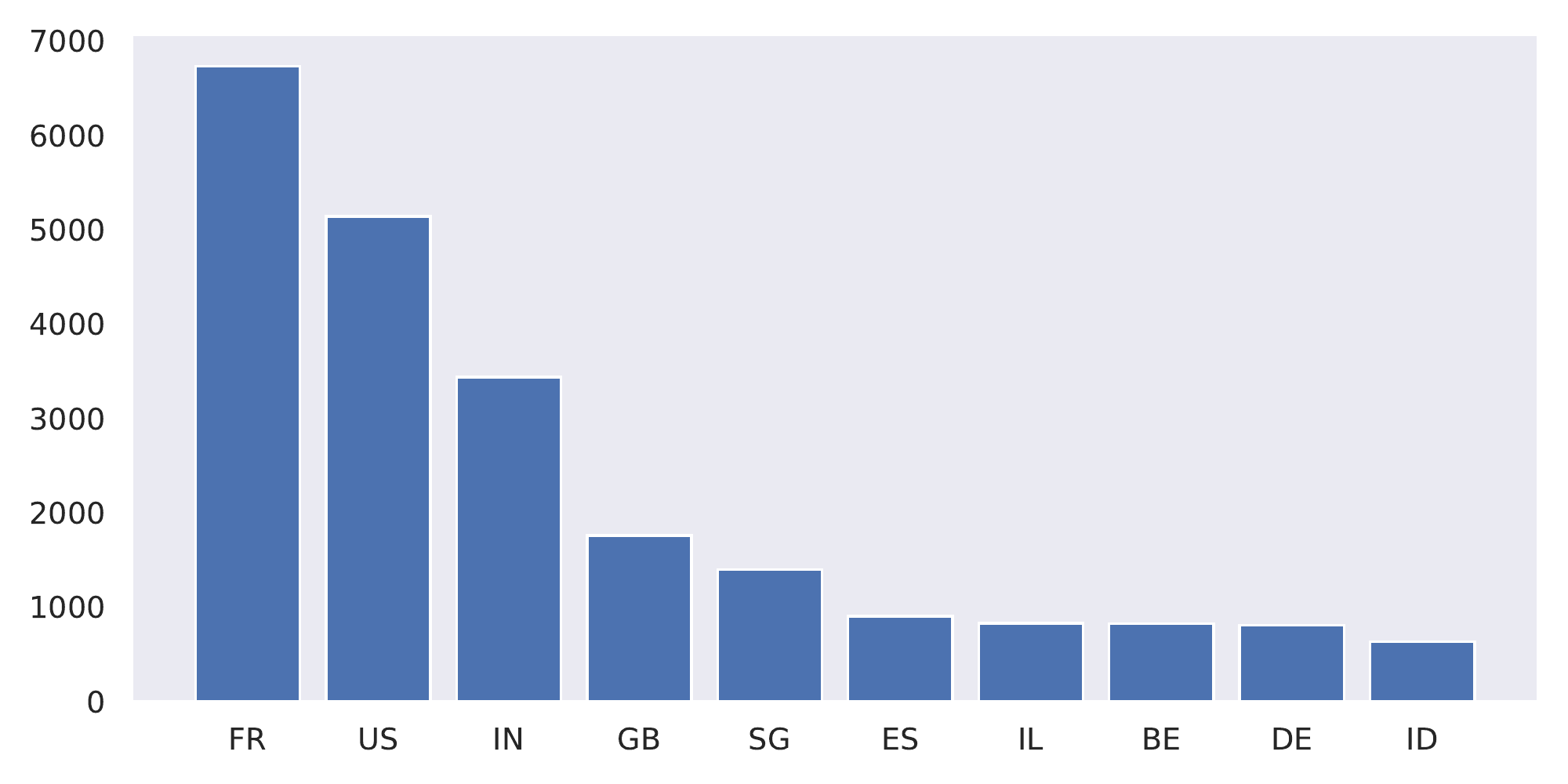}
	\caption{Geographical distribution of the top-10 countries in our dataset.}
	\label{fig:dist_countries}
\end{figure}

Distinguishing feature of our dataset is its geographical variety of the companies. While most of the previous works focus on the startups from the USA\cite{giardino2014early, zhang2017predicting}, we analyze the startups all across the world with a slight focus on Europe.  Figure~\ref{fig:dist_countries} illustrates the distribution of top-10 countries in our dataset.

\subsection{Features}\label{subsec:features}
Once the data from the above sources has been collected, one need to extract a proper set of features in order to define a space where the prediction task can be done efficiently. Below, we describe four categories of features that we extracted, along with the intuitions behind, from the web for the purpose of startup success prediction.

\subsubsection{General features}

The following features are considered as the core information about startups:
\begin{itemize}
	\item Country of origin,
	\item Age,
	\item Number of employees,
	\item Number of offices,
	\item Number of people featured on Team page of startup's website.
\end{itemize}
Importance of these features for the task of fundraising prediction is quite obvious: venture's evolution in different countries varies. Age as well as the number of employees and offices characterize different stages of startup evolution and the properties of fundraising process strongly depend on the stage of the venture. 

The country of the startup's origin is extracted from the address pages on the company website. We use statistical methods to infer which country is the most likely the country of the company. To do that, we employ regular expressions to extract phone numbers (via country codes), then simply look around the phone number, in a fixed window size, to find the country. The country with the most occurrence is then taken as the country of origin and (in case of ties no country is selected).

To infer the age of the startup, we simply use its creation date. Most countries give public access to a registry of all companies, in which one can usually find the creation date. Another heuristic we use is to infer the creation date from the dates of the creation of different media from the company: website and the social media. In case of different createion dates identified by the two previously mentioned sources, the older date is taken as the creation date. According to our observations, in the context of startups, this is in most cases a very good approximation of creation date: in 28\% of cases, it return the correct creation date and in 72\% of cases, it returns a creation date with maximum of two years shift. 

The number of people featured on Team page of startup's website is extracted as follows: usually the team page follows a repeating template containing information about every person (name, role, social media links, picture, etc). We find and extract this repeating template and then use statistical methods to verify that it corresponds to people names, job functions, etc. Finally, information about the number of employees and offices is extracted from the country-specific databases.\textsuperscript{\ref{after_acceptance}}

\begin{table*}[t]
	\caption{Startup features used in this study.}
	\label{tab:feats}
	\small
	\begin{tabular}{|l|l|p{9cm}|l|c|}
		\hline
		Group & Name & Description & Type & Sparse \\
		\hline
		General & Country & Country of a startup's origin & Categorical & x \\
		& Age & Age of a startup & Numeric &  x \\
		& Number of employees & Official number of employees & Numeric & x \\
		& Number of offices & Official number of offices & Numeric & x \\
		& People on team page & Number of people featured on Team page of a startup's website & Numeric & x\\
		\hline
		Financial & N previous rounds & Total number of previously secured funding rounds & Numeric & x \\
		& Last fundraising amount & The amount of money secured in the last funding round & Numeric & x \\
		& Time since fundraising & Days since the last detected funding round & Numeric & x \\
		& Mean fundraising amount & Mean size of previously secured funding rounds & Numeric & x \\
		& Maximal fundraising amount & Size of the biggest previous funding round & Numeric & x \\
		\hline
		Social Networks & Social media accounts & Does a startup have an account in Facebook/Linkedin/Instagram/Youtube or blog on its own website? & Binary & x\\
		& Twitter statistics & Number of tweets and mean/max likes and retweets obtained for each month during the last year & Numeric & x\\
		& Twitter lingual statistics & Modal language of user tweets in each month during the last year & Categorical & x \\ 
		& Twitter hashtags & How many times a startup used each of the most popular 500 hashtags among startups during the last year & Numeric & \checkmark \\
		\hline
		Web presence & Number of relevant results & Number of pages relevant to the startup in the first 10 result from Google search & Numeric & x \\
		& Total results & Total number of results reported by Google & Numeric & x\\
		& Domains & Number of results from each of 500 most popular domains (only top 10 Google search results are analyzed) & Numeric & \checkmark \\
		
		\hline
		
	\end{tabular}
\end{table*}

\subsubsection{Financial features}

History of startup's previous funding rounds is evidently an important factor for predicting future fundraising as different fundraising rounds happen usually with similar patents w.r.t. the previous rounds secured by the startup. The process of detecting funding events for startups is described in Section~\ref{subsec:labeling}. In this work, we propose to extract the following features to summarize the financial history of a startup:
\begin{itemize}
	\item Total number of previously secured funding rounds,
	\item Last fundraising amount,
	\item Mean and maximal amount of previously secured rounds,
	\item Time since the last secured round.
\end{itemize}
%

\subsubsection{Google search results features}
In \cite{sharchilev2018web}, the authors have shown that a highly useful set of features for the task of the startup success prediction can be extracted from crawling of the observable web for the startup presence. For the purpose of extraction of these features, the authors analyze the data from Yandex\footnote{\url{https://yandex.ru}}, a major Russian search engine. For each startup, they count the number of references to the startup's website on the webpages from different domains. This data, however, is not easily accessible by ordinary web users.

Accordingly, in the present study, similar information using more widely available tools has been extracted, in particular Google search API. For each startup, search results with a date within a year preceding the start of \textit{prediction period} have been analyzed. Given a startup name and a date range, a query to Google API is made and irrelevant results were filtered in order to perform the analysis of domains frequencies similar to \cite{sharchilev2018web}. In order to exclude irrelevant results, we check whether the snippet of a search result contains the startup's name or not. Since the purpose of this work is to build models with entirely free tools, we constrain ourselves to the amount of queries available with the Google Cloud Platform Free Tier \footnote{\url{https://cloud.google.com/free/docs/gcp-free-tier}}. Therefore, we obtain only top 10 results for a given startup name and a date range.
The following statistics are then extracted form these search results:
\begin{itemize}
	\item Number of relevant results: we assume that result is related to startup only if a snippet of result contains the startup's name,
	\item Total number of results as reported by Google,
	\item Number of search results from each of 500 domains \emph{popular} domains.
\end{itemize}
For the latter, we simply sort all the domains appearing in the results based on the number of times they contain the name of startups under investigation. We then take the top 500 domains. The intuition is to take the domains which are more likely to talk about startups and, as a result, reduce the amount of noise in our feature space.

\subsubsection{Social networks presence features}
Over the last two decades, the impact of social networks on different social, economic and political processes became remarkable. Given the fact that for a startup it is crucial to reach the potential audience via social media, this category of features can heavily impact the prediction performance. One can note such impacts in the investigations done in the literature such as \cite{zhang2017predicting} which highlighted the importance of social media presence for a crowdfunding success of a startup.
We use here, first, a set of social media features that are binary and indicate whether a startup has an account in several popular social media:
\begin{itemize}
	\item Facebook,
	\item Instagram,
	\item Linkedin,
	\item YouTube,
	\item Twitter,
	\item Blog on the startup own website.
\end{itemize}
This information is extracted with a simple script that searches for social network buttons on a website of a startup.
The second set of features extracted from social media corresponds to statistical information of a startup's website:
\begin{itemize}
	\item Number of people that give reference to their Linkedin account on the team page of the startup,
	\item Number of entries in blog during the last year.
\end{itemize} 
These two features indicate the willingness of the startup to appear in the social media and to be visible and followed by others.

Because of the important presence of startups on Twitter and since information from Twitter is readily available (contrary to other social media like Facebook and LinkeIn), we also extract features that describe the activity of each startup on this particular social media in the year the precedes the year for which funding events are predicted:
\begin{itemize}
	\item Aggregated monthly startup's Twitter account statistics for the last year including: number of posted tweets, mean/max number of likes and retweets of user's tweets; modal language of user tweets,
	\item total number of different users that mention startup's account in their tweets during the last year,
	\item Information about \textit{hashtags} used by the startup: \emph{a)} for all startups, we collect their last 2300 tweets from which we establish a list of the 500 most frequently used hastags; each startup is then represented as a 500-dimensional vector the dimensions of which correspond to the number of times the startup used the hashtag during the last year.
\end{itemize}

Table~\ref{tab:feats} summarizes the features explained above. The type of each feature (categorical or numerical) as well as its nature (sparse or dense) are also illustrated in the two rightmost columns.

%

\subsection{Data labeling}\label{subsec:labeling}
Another challenge in solving the task of predicting startup success from open sources is labeling the data. While commercial databases often contain dates of funding events where amounts are usually extracted manually by human experts, we aim to automatically detect startups fundraising from news and Twitter. For this purpose, we subscribed to RSS feeds for a large set of news websites focused on startups.\footnote{Some examples are given in Appendix~\ref{app:data_label}. The exhaustive list countains several thousands of entries.}

For each sentence in tweets and news headlines, we proceed as follows:

\begin{enumerate}
	\item identify money amounts using regular expressions (ex: \$5M).
	\item identify mentions of known startup using either startup names or Twitter screen names.
	\item retain as a candidate funding event the startup mention and money amount if they are separate as a fundraising verbs\footnote{The complete list: raise, take, get, grab, score, secure, receive, close, announce, complete}.
	\item merge candidate funding events if they occur within a three-month period as given by the tweet or news date.
\end{enumerate}

\begin{figure}[t]
	\includegraphics[width=\columnwidth]{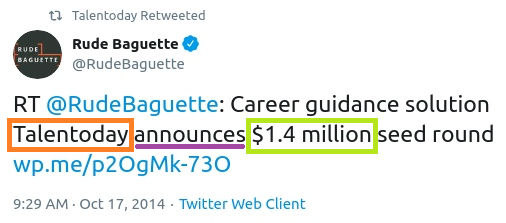}
	\caption{Illustration of the funding extraction algorithm: detected money amount and startup name are shown in boxes; fundraising verb is underlined.}\label{fig:label_detection}
\end{figure}

An example of this approach is illustrated in Figure \ref{fig:label_detection} where the verb is underlined in purple, and startup name and the amount are shown in orange and  green boxes respectively.

This algorithm, despite its simplicity, is able to extract a surprisingly large amount of funding events. Indeed, we were able to detect 9139 funding events that took place during the last 5 years corresponding to an average of about 7\% startups securing a funding event each year.

This said, the above algorithm may falsely assign funding events to startups. For example, in particular when there are several startups with similar names. In order to estimate the error rate in the detected funding events, we sampled 200 funding events detected by our algorithm and manually checked if they were correct or not. We found 17 false funding events, \emph{i.e.} meaning that the rate of false positives amounts to 8.5\%.

Another problem is that not all funding events are identified as funding events can be reported in different ways. In addition, it is worth noticing that information about some of the funding rounds is held private and, therefore, cannot be extracted. In order to estimate the number of funding events not identified by our algorithm, we randomly sampled 200 startups for which we did not find a funding event in a chosen one-year time period and found that 12 of them (\emph{i.e.} 6\%) actually raised money.

%% file: 4Prediction.tex
\section{Prediction models}\label{sec:models}
In this section, we describe the machine learning frameworks we explored to predict new funding events. 
\subsection{Positive-Unlabeled setting}\label{subsec:pumodel}
The procedure of data labeling that we proposed in \ref{subsec:labeling} does not include assigning negative labels to the startups. Thus, it might be natural to consider our problem in the context of Positive-Unlabeled (PU) learning\cite{Bekker2018LearningFP}. 

PU learning is a field of machine learning that studies the principles of training a binary classifier in a case when some positive objects are labeled as positive while the majority of objects does not have any label and consists of objects of both classes. The PU learning methods can generally be divided into two categories. The first one aims to find true negatives in the data and then perform normal binary classification \cite{Liu2002PartiallySC} \cite{Li2003LearningTC}. The second category aims to modify the loss function to account for the PU setting. \cite{elkan2008learning} first proposed to treat unlabeled data as weighted positive and negative data mixture. \cite{du2014analysis} introduced the approach described below.

Let $X \in \mathbb{R}^d$ and $Y \in \{\pm 1\} (d\in \mathbb{N})$ denote the input and output random variables, $p_p(x) = p(x| Y = +1)$ and $p_n(x) = p(x| Y = -1)$ be the positive and negative marginals, $\pi_p=p(Y=+1)$ and $\pi_n=p(Y=-1)=1-\pi_p$ stand for class-prior probabilities.

For classification task the empirical risk $\hat{R}$ is minimized, which can be decomposed into empirical risk on positive and negative examples: 
\begin{equation}
\hat{R} = \pi_p \hat{R}_p^- + \pi_n \hat{R}_p^+
\end{equation}

Given the loss function $l$ and the decision function $d$ one can estimate $\hat{R}_p^+ = (1/n_p) \sum_{i=0}^{n_p} l(g(x^p_i), +1)$. The difficulty of PU learning is the lack of data sampled from $p_n(x)=p(x|Y=-1)$, therefore it is not possible to use estimate  $\hat{R}_n^- = (1/n_n) \sum_{i=0}^{n_n} l(g(x^n_i), -1)$. However, the following approximation can be used: 

\begin{eqnarray}\label{px}
p(x) &=& \pi_n p_n(x) + \pi_p p_p(x) \\
\pi_n p_n(x) &=& p(x) - \pi_p p_p(x)
\end{eqnarray}

\begin{dmath}
	\pi_n \hat{R}_n^- = \hat{R}^-_U - \pi_p \hat{R}_p^- = (1/n_u) \sum_{i=1}^{n_u} l(g(x^u_i), - 1) - \pi_p (1/n_p)\sum_{i=1}^{n_p}l(g(x_i^p), -1) 
\end{dmath}

\begin{dmath}
	\hat{R}(g) = \pi_p \hat{R}_p^+ + \hat{R}^-_U - \pi_p \hat{R}_p^-=\pi_p  (1/n_p)\sum_{i=1}^{n_p}l(g(x_i^p), +1) + (1/n_u) \sum_{i=1}^{n_u} l(g(x^u_i), - 1) - \pi_p (1/n_p)\sum_{i=1}^{n_p}l(g(x_i^p), -1)
\end{dmath}

In \cite{Kiryo2017PositiveUnlabeledLW} the authors propose to modify the loss to make it more suitable to neural networks. They note that the theoretical guaranties obtained for the method in \cite{du2014analysis} need an assumption on the Rademacher complexity\cite{Mohri:2012:FML:2371238}, which does not hold for the neural networks. At the same time they show that in practice neural networks overfit while trying to make the term $\pi_p (1/n_p)\sum_{i=1}^{n_p}l(g(x_i^p), -1)$ as large as possible to get negative loss. They propose simple modification of loss to mitigate the problem:
\begin{dmath}\label{nnPu}
	\hat{R}(g) = \pi_p \hat{R}_p^+ + \max(\hat{R}^-_U - \pi_p \hat{R}_p^-, 0)
\end{dmath}

In our work we apply this modification called Positive-Unlabeled Learning with Non-Negative Risk Estimator(nnPU) in order to see whether it is beneficial compared to the standard binary classification. To this end we construct a multilayer perceptron and train it on our dataset \emph{a)} with the standard sigmoid loss and \emph{b)} with the loss proposed in \cite{Kiryo2017PositiveUnlabeledLW}.

Another approach to PU learning that we applied to our data is Positive Unlabeled AUC Optimization \cite{Sakai2017SemisupervisedAO}. In the standard binary classification AUC for a classifier $g$ is defined as:

\begin{equation}
AUC(g) = \mathbb{E}_p [\mathbb{E}_n[I(g(x^P)\ge g(x^N))]]
\end{equation}

where $\mathbb{E}_p$ and $\mathbb{E}_n$ are the expectations over $p_p(x)$ and $p_n(x)$, respectively. $I(·)$ stands for the indicator function.

In practice a composite classifier $f(x,x') = g(x) - g(x')$ can be trained by minimizing the empirical AUC risk \cite{herschtal2004optimising} \cite{davis2006relationship} defined as:

\begin{equation}
\hat{R}(f) = \frac{1}{n_P n_N} \sum_{i=1}^{n_P} \sum_{j=1}^{n_N} l(f(x_i^P, x_j^N))
\end{equation} 

where $l(m)$ is a surrogate loss.

Using formula \ref{px} in \cite{Sakai2017SemisupervisedAO} the authors propose the following expression for AUC risk in PU setting:

\begin{dmath}
\hat{R}_{PU}(f) = \frac{1}{\pi_n n_P n_U} \sum_{i=1}^{n_P} \sum_{j=1}^{n_U} l(f(x_i^P, x_j^U)) - \frac{1}{\pi_n n_P (n_P - 1)} \sum_{i=1}^{n_P} \sum_{i'=1}^{n_P} l(f(x_i^P, x_{i'}^P)) + \frac{\pi_p}{\pi_n (n_p - 1)}
\end{dmath}

and further construct a kernel based method that efficiently optimizes the given PU risk. In our work we use the authors original implementation of this method in python available on GitHub \footnote{\url{https://github.com/t-sakai-kure/pywsl}} to solve the task of startup success prediction.



\subsection{Positive-Negative setting}\label{subsec:PN_setting}
Despite the arguments in favor of PU setting for our task, there are also several factors that incline us to give preference to the traditional binary classification setup, referred to as PN for Positive-Negative(PN), which consists in treating the startups for which we could not detect funding events as negative examples. The most important one is that currently the PU setting is rather restrictive in terms of available algorithms. The proposed modifications of the loss function make the loss either non-convex \cite{du2014analysis} \cite{Kiryo2017PositiveUnlabeledLW} or non smooth \cite{Plessis2015ConvexFF} which impedes their use with the algorithms relying on the second order optimization techniques. Another point is that the theoretical results for the PU learning are obtained with assumption that the objects labeled as positives are always positive. At the same time in our dataset, as discussed in \ref{subsec:labeling} the share of falsely assigned funding events is estimated to be around 9\%. The last consideration is that due to the class imbalance in our dataset, treating all the unlabeled objects as negative might be less harmful than in class-balanced PU setting.

For the given reasons we performed a series of experiments in traditional binary classification settings. We tested the performance of the most widely used machine learning models such as Logistic Regression, Random Forest, and a recent gradient boosting algorithm with support of categorical variables called CatBoost \cite{prokhorenkova2018catboost}. In the preliminary experiments, we also compared different popular gradient boosting algorithms CatBoost \cite{prokhorenkova2018catboost}, XGBoost \cite{chen2016xgboost} and LightGBM \cite{ke2017lightgbm} and found that they yield similar performance on our dataset. Therefore, we only report the results obtained with CatBoost. As discussed in \ref{subsec:pumodel} in order to study the impact of nnPU loss modification we also trained a neural network in traditional binary classification setting. 
%
%
We also did our best to reproduce the approach WBSSP introduced in \cite{sharchilev2018web}. Since this algorithm is very specific to the dataset for which it was developed \emph{i.e.} it specifies which exact feature should go to which logistic regression group, it is impossible to exactly reproduce it for a dataset with a very different set of features. However, we followed the general idea and built logistic regression models on semantic groups of features, and then built a CatBoost model using logistic regressions as features in addition to non-sparse initial features of the dataset. The details for about semantic groups of features as well as the information about sparsity can be found in table \ref{tab:feats}.

%% file: 5Evaluation.tex
\section{Evaluation}\label{sec:eval}

In this section, we detail the experiments conducted to assess the effectiveness of the proposed model. We also investigate the effectiveness of the PU framework  (see Section~\ref{sec:models}), in addition to an extensive discussion on the importance of (set of) features.

\subsection{Data split and metrics}

Our algorithm of populating train and test sets is identical to the one described in \cite{sharchilev2018web}: we design a function that, given a name of a startup and a date $d$, extracts the feature vector $X$ of the startup using only the information available before the date $d$ (\emph{	e.g.} previous funding events, earlier activity on social networks etc.). Another function, given a name of a startup and a date $d$, returns the binary label -- whether a funding event was detected for the startup in a year since the date $d$. For $d \in$ \{01-09-2014, 01-09-2015, 01-09-2016, 01-09-2017 \}, we extract $(X, y)$ pairs for each startup in the list and populate train set. For $d = \text{01-09-2018}$ the extracted pairs go to test set.  


To evaluate our results, we use the area under the ROC curve (AUC) \cite{fawcett2006introduction}, which illustrates the behavior of the prediction w.r.t. True Positive Rate (TPR) and False Positive Rate (FPR) at different points, and has been used in a vast variety of tasks to assess the classification performance \cite{hand2001simple, korolev2017residual}. It furthermore can properly assess the effectiveness of classification models in the presence of noisy labels as well as in the case of imbalanced classes \cite{zhang2014active, he2009learning}.
On top of that, we adopt the same strategy as \cite{sharchilev2018web} and used the F-score with $\beta=0.1$ in order to have a more significant impact of precision and, more importantly, to be able to compare with the results presented in \cite{sharchilev2018web}. We also assess the performance of the models via precision on the top 100 ($P@100$) and on the top 200 ($P@200$) results.

As discussed in Section~\ref{subsec:labeling}, label extracted by our method are sometimes incorrect. In \cite{jain2017recovering}, the authors proposed a way to estimate binary classifier performance in PU setting. They theoretically show, that if $f_1$ and $f_2$ are the distributions of the positive and negative objects, the unlabeled examples are drawn from the distribution $f(x) = \alpha f_1(x) + (1-\alpha)f_0(x)$ and the labeled examples come from the distribution $g(x) = \beta f_1(x) + (1-\beta) f_0(x)$ the true AUC of a classifier can be recovered from the obtained on the noisy labels AUC$^{pu}$ with the following formula:

\begin{equation}\label{auc_correction}
	\text{AUC} = \frac{\text{AUC}^{pu} - \frac{1-(\beta - \alpha)}{2}}{\beta - \alpha}
\end{equation}

where $\alpha$ and $\beta$ are related to the amount of noise in the labels. As described in Section~\ref{subsec:labeling}, for our dataset, we experimentally estimate $\alpha=0.06$, since 6\% of the startups, for which we did not identify funding events, \textit{i.e.} unlabeled startups actually received money, and $\beta= 1-0.085 = 0.915$, since there exists 8.5\% of falsely detected funding events. Therefore, for all our experiments, we report both raw value of AUC and the AUC value corrected according to the Equation~(\ref{auc_correction}). 

\subsection{Positive-Unlabeled results}

As discussed in Section~\ref{subsec:pumodel}, we conducted some experiments in order to find out whether the PU learning can be beneficial compared to the traditional binary classification setting or not. To this end, we constructed a multilayer perceptron (MLP) model with two hidden layers of size 100 and had it trained with learning rate 0.0001 and batch size of 1000 for 5 epochs. As mentioned in Section~\ref{subsec:pumodel}, we first run the network with a standard sigmoid ($\sigma$) loss function, and then with the PU modified loss function (Non-Negative Risk Estimator) that has been presented in \cite{Kiryo2017PositiveUnlabeledLW}, a method we refer to as nnPU. Finally we investigated the kernel-based approach explained in \cite{Sakai2017SemisupervisedAO}, referred to as PU-AUC hereafter.

Table~\ref{tab:PU_results} illustrates the results of these experiments. The upper part shows the two neural network methods while the last line is separated from the rest as the objective function is different. The first observation one can make is that among the neural network methods, the one based on PU yields better results. Indeed, according to a Wilcoxon rank test, it is significantly better with $p < 0.05$ for  $P@200$ and $F@200$, and with $p<0.01$ for $ROC-AUC$.
\begin{table}[t]
	\caption{PU setting. The upper part of the table illustrated the neural network approaches and the lower part shows the direct optimization of AUC from \cite{Sakai2017SemisupervisedAO}. Neural networks are trained with 10 random seeds, mean and std. reported.}
	\label{tab:PU_results}
	\footnotesize
	\begin{tabular}{|l|c|c|c|c|p{1.3cm}|}
		\hline
		Loss function&  $P@100$ &  $F_{0.1}@100$ &  $P@200$ & $F_{0.1}@200$ &  ROC-AUC raw/corrected \\
		\hline
		Standard $\sigma$  & 0.26(0.02)  & 0.22(0.01) & 0.24(0.01) & 0.22(0.01) & 0.78(0.01) / 0.83(0.01) \\
		nnPU & 0.27(0.01) & 0.22(0.01) & 0.25(0.01) & 0.23(0.01) & 0.79(0.01) / 0.84(0.01) \\ \hline \hline
		PU-AUC & \textbf{0.45} & \textbf{0.38} & \textbf{0.43} & \textbf{0.39} & \textbf{0.82/0.87} \\ \hline 
	\end{tabular}
\end{table}
In addition, the best results are obtained with PU-AUC that significantly outperforms all the other methods on all metrics at $p<0.01$.
It is however hard to say whether the difference should be attributed to the modification of the loss function for imbalanced dataset or to the different nature of the base algorithms. All in all, these results show that PU based approaches yield better results than a simple PN method as MLP (or even Logistic Regression as illustrated below).

\subsection{Positive-Negative results}\label{subsec:PN_results}

For traditional binary classification setting, we use implementations of Logistic Regression(LR) and Random Forest(RF) from Scikit-learn \cite{scikit-learn} and the recent gradient boosted tree method CatBoost \cite{prokhorenkova2018catboost}. For all the methods, we perform 5-fold cross validation of hyperparameters on the train set. For the CatBoost, we found that using rather a high value of  the coefficient at the $L_2$-regularization term of the cost function 100 helps to mitigate the overfitting problem.

To implement our version of WBSSP \cite{sharchilev2018web}, we used the same Logistic Regression and CatBoost implementations. For each group illustrated in Table \ref{tab:feats}, we built a Logistic Regression; the training set is split into 5 folds, and out-of-fold predictions are used for the downstream classifier. At the same time, features that do not have sparse flag were directly fed into the final CatBoost classifier.  


Table~\ref{tab:PN_results} shows the results of the three employed ensemble methods as well as the WBSSP-based approach, \emph{i.e.} the one explained above. As mentioned in Section~\ref{subsec:PN_setting}, this approach is particularly designed for the dataset presented in \cite{sharchilev2018web} and, as a result, it cannot be compared directly to our approach. However, it is the closest pipeline to that of \cite{sharchilev2018web} which adapts to our dataset.

\begin{table*}[ht]
	\caption{PN setting models comparison. The best values are shown in bold. Values shown in parenthesis and marked with *  are calculated on the labels corrected by the human experts for the top 200 companies.}
	\label{tab:PN_results}
	\begin{tabular}{|l|c|c|c|c|c|c|}
		\hline
		Model &  $P@100$ &  $F_{0.1}@100$ &  $P@200$ &  $F_{0.1}@200$ &  ROC-AUC raw & ROC-AUC corrected\\
		\hline
		LR &  0.380 & 0.317 & 0.345 & 0.315 & 0.774 & 0.821\\
		RF & \textbf{0.580} & \textbf{0.483} & 0.470 & 0.429 & 0.796 & 0.847\\
		CatBoost & 0.53 (0.640*) & 0.442(0.531*) & \textbf{0.480 (0.580*)} & \textbf{0.439(0.528*)} & \textbf{0.834} & \textbf{0.890} \\ 
		WBSSP-based & 0.52 & 0.433 & 0.470 & 0.429 & 0.825 & 0.881\\
		\hline
	\end{tabular}
\end{table*} 

As it can be seen in Table~\ref{tab:PN_results}, Logistic Regression totally fails to provide good results with respect to the two other ensemble methods, \emph{i.e.} RF and CatBoost. That can be simply explained by the fact that Logistic Regression is not able to predict the loss from the designed features which are difficult to be separated linearly. This basically comes form the complexity and the heterogeneous nature of features explained in Section~\ref{sec:framework}.

Ensemble methods, however, are able to overcome this complexity and stress on important features in order to linearly separate subregions of the space and combine them, via weak learners, in order to perform better predictions. If we compare however RF and CatBoost, we can see that differences between top-100/200 precision and F$_{0.1}$ scores are small. CatBoost is nevertheless significantly better than RF in terms of ROC-AUC. This can be explained by the fact that CatBoost benefits from the gradient boosting framework and is able to perform optimization in functional space. On top of that, compared to the RF, it behaves in a more robust way in dealing with categorical and heterogeneous features, as it benefits from history-based ordered target statistics \cite{prokhorenkova2018catboost}. All in all, CatBoost is the best performing model over all PN and PN methods that we investigated.

After selecting the best model, based on its significantly better ROC-AUC, we set out to obtain the estimate of its performance not contaminated by the noise in our test set labels. To this end, we took the list of top-200 companies according to our model and asked a human expert to manually check label for each startup. The metrics calculated with the corrected labels are shown in parentheses in Table~\ref{tab:PN_results}. 


As our last analysis on these experiments, we illustrate in Table~\ref{tab:with_yandex} the best results of Table~\ref{tab:PN_results}, \emph{i.e.} CatBoost, along with those reported in \cite{sharchilev2018web}. Although there is no direct way to compare metrics obtained on two completely different datasets, the categories of features are rather similar and, accordingly, can provide a good insight into these two approaches. In terms of size and class balance the datasets are also comparable: 15128 objects with 8\% of positives in \cite{sharchilev2018web} dataset vs. 33165 objects with 6.3\% of positives in our dataset. As it can be seen from the table, the results reported by the same metrics illustrate that our classifier scores are consistently higher than the ones reported in \cite{sharchilev2018web}.

\begin{table}[h]
	\caption{Comparison to the results reported in \cite{sharchilev2018web} on the dataset presented therein with WBSSP pipeline. Values marked with *  are calculated on the labels corrected by the human experts for the top 200 companies.}
	\label{tab:with_yandex}
	\footnotesize
	\begin{tabular}{|p{1.48cm}|c|c|c|c|p{1.32cm}|}
		\hline
		Model &  $P@100$ &  $F_{0.1}@100$ &  $P@200$ &  $F_{0.1}@200$ &  ROC-AUC \\
		\hline
		Our best model & 0.640*& 0.531* & 0.580* & 0.528* & 0.890 \tiny{(corrected)} \\ 
		WBSSP \cite{sharchilev2018web} & 0.626 & 0.383 & 0.535 & 0.439 & 0.854 \\
		\hline
	\end{tabular}
\end{table}
%



\subsection{Ablation Analysis}
In studies like the present one, the diversity of possible numerical and categorical feature makes it sometimes difficult to have deep insights on the prediction models. Accordingly, a feature importance analysis is usually crucial in order to better understand and analyze the model. For this reason, we performed an ablation analysis aiming at assessing the importance of the different feature groups.

To this end, we repetitively exclude one semantic groups of features, presented in Table~\ref{tab:feats} in Section~\ref{sec:framework}, and measure how it affects the model performance. Evidently, we keep always the General category as it contains the core information needed for the model.
Table~\ref{tab:ablation} presents the results of this analysis. 
The fist observation one can make from this table is that including all the features provides the best performance regardless of the metric. This is an important point as it indicates that all the categories presented in  Section~\ref{sec:framework} are involved in boosting the performance of the prediction task. The second point is that social network information plays an important role in boosting the performance as removing it brings the most important deterioration to all metrics but ROC-AUC. When it comes to overall performance, measured by ROC-AUC, the removal of each semantic group negatively impacts the performance, again indicating that all features are important to predict funding events.



\begin{table}[t]
	\caption{Ablation Analysis}
	\label{tab:ablation}
	\footnotesize
	\begin{tabular}{|p{1.7cm}|c|c|c|c|p{1.2cm}|}
		\hline
		Features &  $P@100$ &  $F_{0.1}@100$ &  $P@200$ &  $F_{0.1}@200$ &  ROC-AUC \scriptsize{raw/corrected} \\
		\hline
		All features & 0.530 & 0.442 & 0.480 & 0.439 & 0.834/0.890 \\
		No Financial & 0.530 & 0.442 & 0.440 & 0.402 & 0.819/0.873 \\
		No Social Net. & 0.480 & 0.400 & 0.405  & 0.370 & 0.820/0.873 \\
		No Web presence & 0.500 & 0.417 & 0.475 & 0.434 & 0.808/0.861 \\
		\hline
	\end{tabular}
\end{table}

\subsection{Feature Importance}

To complete the analysis of the features retained, we make use of a recent approach particularly designed for interpreting the results of tree-based ensemble methods. This technique, abbreviated as SHAP for SHapley Additive exPlanations, was introduced in \cite{DBLP:journals/corr/abs-1802-03888} and aims at addressing the inconsistency in standard feature importance scoring methods used in, \emph{e.g.}, 
 \cite{irrthum2010inferring, Chen2016XGBoostAS, auret2011empirical}.

SHAP provides, for each feature and each example, a measure of the impact of the feature on the decision that a model makes on the example. The calculation of this impact is based on the comparison of classifier's output on a full feature vector and the expectation of the classifier's output over feature vectors with the studied feature value replaced by all the possible values of the feature. SHAP plots are then constructed, where the $x$-axis corresponds to SHAP impact values and the $y$-axis to the different features. A dot on the figure finally corresponds to an example for which the corresponding feature ($y$-axis) has the SHAP impact value given in the $x$-axis. Note that on the $y$-axis, features are sorted according to their importance, the topmost feature being the most important one. The importance of a feature is measured by given by the sum, over all examples, of the absolute values of the SHAP impact scores.

\begin{figure*}

\subfloat[Financial features]{%
	\includegraphics[width=0.99\textwidth]{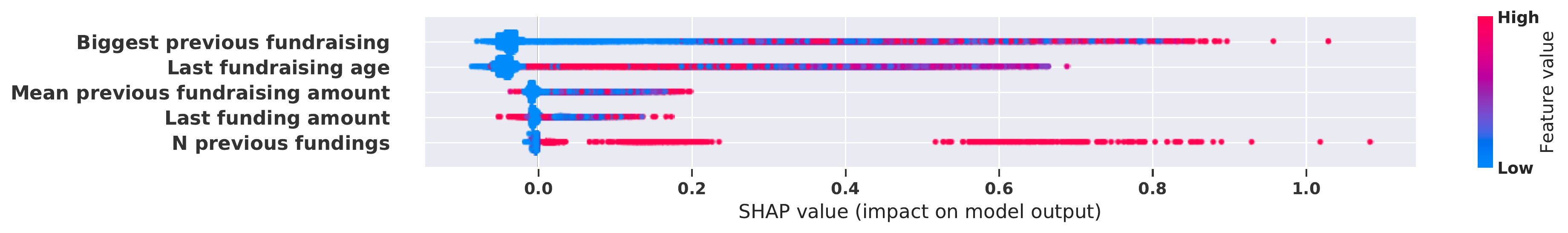}
}

\subfloat[Web presence]{%
	\includegraphics[width=0.99\textwidth]{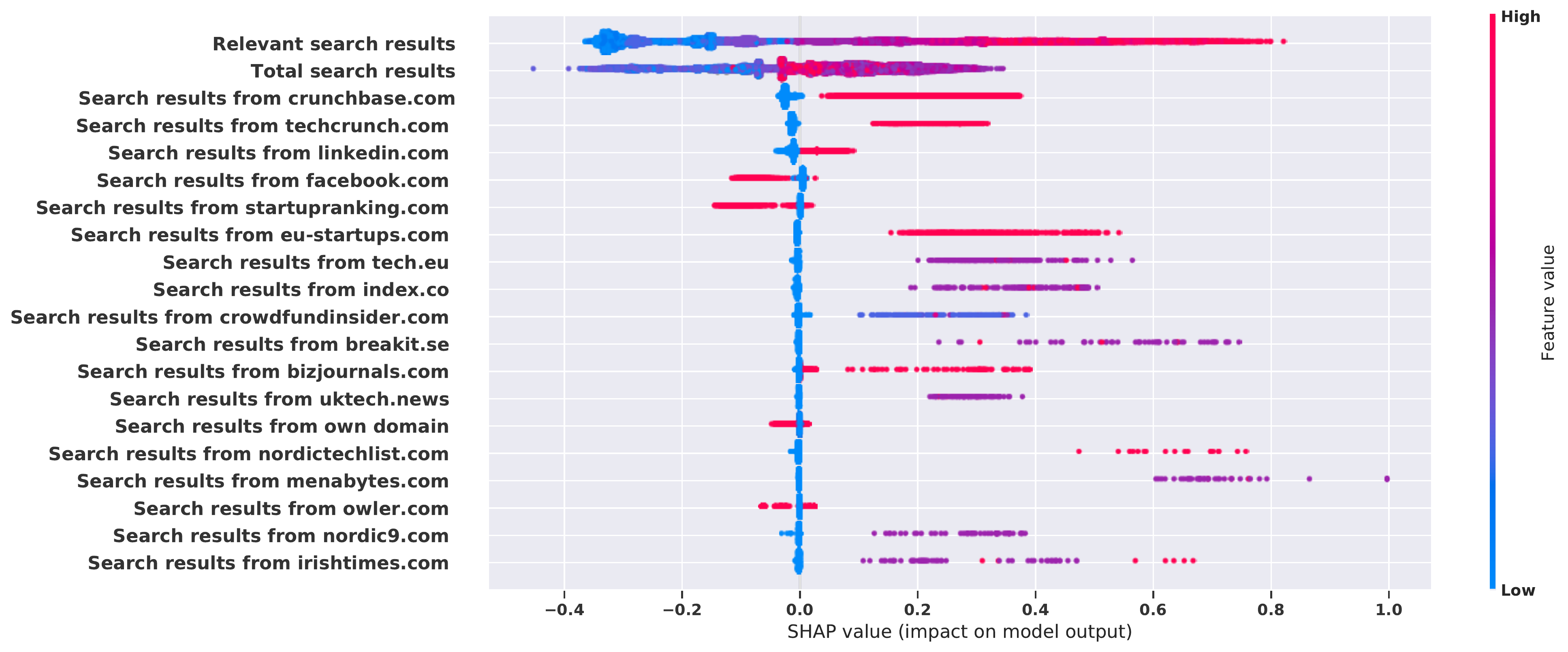}
}

\subfloat[Social media activity]{%
	\includegraphics[width=0.99\textwidth]{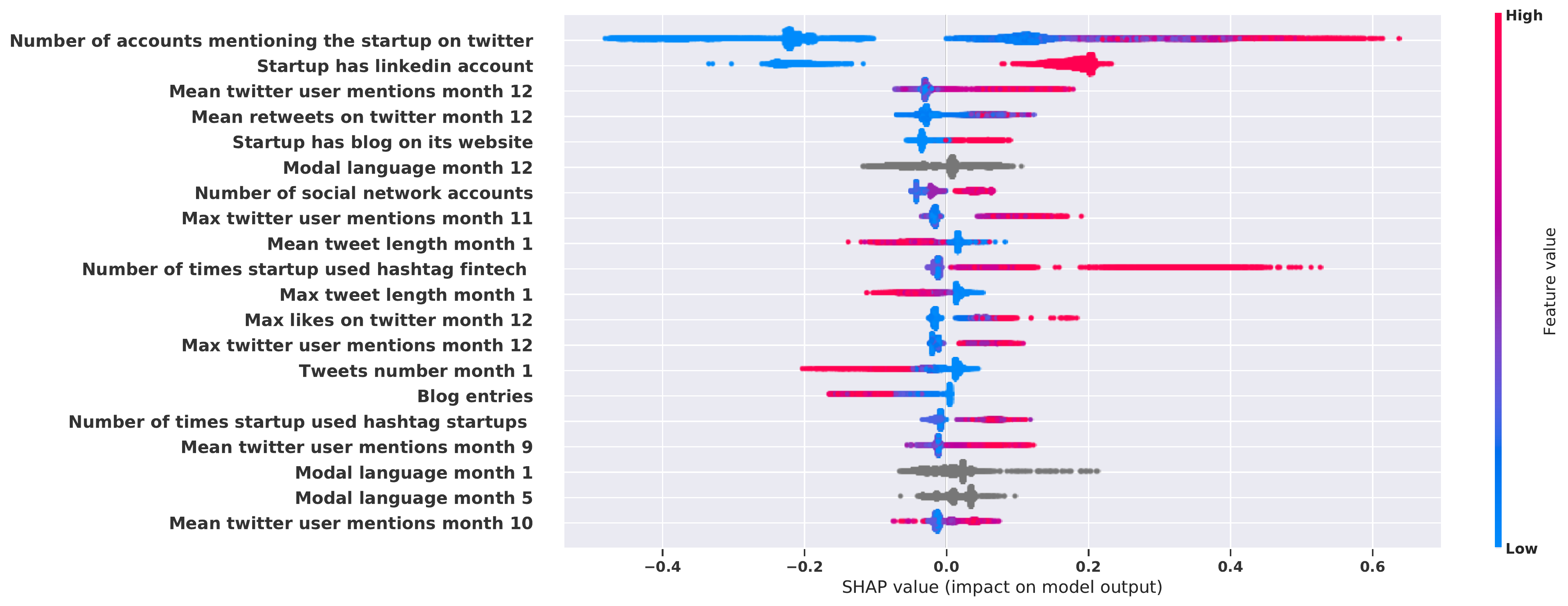}
}

\caption{Feature importance analysis using SHAP-values for the different semantic groups of features (a) financial, (b) the web presence and (c) social media activities. Features are stacked vertically based on their importance from top to bottom of each figure. Each dot represent an instance in the dataset with its corresponding SHAP value on X-axis.}
\label{fig:shap}

\end{figure*}

Figure~\ref{fig:shap} displays the SHAP plots we obtained for the different semantic groups. Note that all the features are displayed for the financial group, whereas only the 20 most important features are displayed for the web-presence and the social network groups. Several conclusions can be drawn from Figure~\ref{fig:shap}. We present here the most obvious ones. First, the amount of the pages mentioning the startup among the first 10 results provided by a search engine for a query that contains the startup's name is an important feature for predicting the future fundraisings. The same can be said about the number of different people that mention the startup on Twitter. Second, if the pages on LinkedIn or Crunchbase returned by search engine are a positive indicator for future fundings, the pages returned from Facebook or startupranking.com are a negative indicator. Third, not only users mentioning a startup on twitter are a positive sign for the future funding rounds, but also a startup mentioning other users is. 

%

%% file: 6Conclusion.tex
\section{Conclusion}\label{sec:conclu}

We have studied in this paper the problem of predicting funding events for startups. To do so, and contrary to previous studies that have used information from commercial databases, we have solely relied on information that can be extracted from freely, publicly available sources as startup websites, social media and company registries. The features we rely on can be easily obtained from these sources. Furthermore, the prediction models we use are simple and wide-spread; ensemble methods in a standard positive-negative setting indeed yield the best prediction results. Despite these constraints (easily obtained features, simple prediction models), that guarantee that our methods can be re-implemented, the results we obtain are better than the ones obtained with more information sources and more complex models.

Several aspects of our work can nevertheless be improved. First of all, we want to further explore the importance of each feature and explain why certain features (as the presence on specific social networks) are negative indicators for predicting funding events. Second, we plan on using parsing techniques to limit the  amount of false positives when detecting funding events in tweets and news headlines. We also plan to use word embeddings to identify additional \textit{fundraising} verbs, and more generally operators, in order to increase the recall (yet, the most important problem to solve is the one related to false positives). In addition, several sources of information, also readily available, could be envisaged to complement the features we have considered. Patent databases could for example be mined in order to get indicators of the invention portfolio of a startup, an element that is taken into account by many investors. Publicly available information from investment companies, from which technological domain and market information can be inferred, could also be used to further predict which investor is likely to be interested in which startup. The space of potential use cases is large and we hope that the current work will pave the way to new studies on startup analysis.

%% file: Appendix.tex
\appendix
\section{Business register sources}\label{app:biz_reg}
\begin{itemize}
	\item \url{https://businessregister.kompany.com}
	\item \url{https://portal.kyckr.com/}
	\item \url{https://data.opendatasoft.com/explore/dataset/sirene%40public/api/}
	\item \url{https://www.societe.com/}
	\item \url{https://www.infogreffe.fr/}
	\item \url{https://developer.companieshouse.gov.uk/api/docs/}
	\item \url{https://beta.companieshouse.gov.uk/}
	\item \url{https://eng.kurzy.cz/prodej-dat/databaze-firmy.htm}
	\item \url{https://www.ytj.fi/en/index/whatisbis/opendata.html}
	\item \url{https://www.ytj.fi/en/}
	\item \url{http://avoindata.prh.fi/tr_en.html#/}
	\item \url{http://kbopub.economie.fgov.be/kbopub/zoeknaamfonetischform.html}
	\item \url{https://kbopub.economie.fgov.be/kbo-open-data/login?lang=fr}
	\item \url{https://search.cro.ie/company/CompanySearch.aspx}
	\item \url{https://services.cro.ie/overview.aspx}
	\item \url{https://datacvr.virk.dk}
	\item \url{https://www.brreg.no/home/}
	\item \url{https://www.unternehmensregister.de/ureg/?submitaction=language&language=en}
	\item \url{https://www.registroimprese.it/en/}
	\item \url{http://www.rmc.es/Sociedades.aspx}
	\item \url{http://www.infocif.es/}
	\item \url{http://www.fi.se/en/our-registers/company-register/}
	\item \url{https://www.zefix.ch/fr}
	\item \url{https://firmenbuch.at/}
	\item \url{https://www.rcsl.lu/mjrcs/}
	\item \url{https://companies-register.companiesoffice.govt.nz/}
	\item \url{https://ica.justice.gov.il/GenericCorporarionInfo/SearchCorporation?unit=8}
	\item \url{https://beta.registresdentreprisesaucanada.ca/chercher}
	\item \url{https://www.registreentreprises.gouv.qc.ca/RQAnonymeGR/GR/GR03/GR03A2_19A_PIU_RechEnt_PC/PageRechSimple.aspx}
	\item \url{https://www.sec.gov/edgar/searchedgar/companysearch.html}
	\item \url{http://developer.edgar-online.com/apps/mykeys}
	\item \url{https://icis.corp.delaware.gov/Ecorp/EntitySearch/NameSearch.aspx}
	\item \url{http://www.gsxt.gov.cn/index.html}
	\item \url{https://www.gov.ph/data/search/type/dataset}
	\item \url{https://data.gov.sg/}
	\item \url{http://www.ocr.gov.np/index.php/np/}
	\item \url{https://data.gov.in/catalog/company-master-data}
	\item \url{http://seninfogreffe.com/}
\end{itemize}

\section{Some examples of sources for data labeling}\label{app:data_label}
\begin{itemize}
	\item \url{https://500.co/feed/}
	\item \url{https://agfundernews.com/feed/}
	\item \url{http://www.arabianbusiness.com/feed/startup/feed.xml}
	\item \url{https://www.austrianstartups.com/feed/}
	\item \url{https://betakit.com/feed/}
	\item \url{https://bothsidesofthetable.com/feed}
	\item \url{https://www.businessweekly.co.uk/rss.xml}
	\item \url{https://www.cnbc.com/id/10001274/device/rss}
	\item \url{http://www.ecap-partner.com/news/feed/}
	\item \url{https://www.entrepreneur.com/latest.rss}
	\item \url{http://www.finsmes.com/feed}
	\item \url{https://www.frenchweb.fr/feed}
	\item \url{https://www.geekwire.com/startups/feed/}
	\item \url{https://iamanentrepreneur.in/feed/}
	\item \url{https://tech.economictimes.indiatimes.com/rss/startups}
	\item \url{http://knowstartup.com/feed/}
	\item \url{https://www.entrepreneur.com/latest.rss}
	\item \url{http://www.finsmes.com/feed}
	\item \url{https://www.frenchweb.fr/feed}
	\item \url{https://www.geekwire.com/startups/feed/}
	\item \url{https://iamanentrepreneur.in/feed/}
	\item \url{https://tech.economictimes.indiatimes.com/rss/startups}
	\item \url{http://knowstartup.com/feed/}
	\item \url{https://www.maddyness.com/feed}
	\item \url{http://feeds.mashable.com/Mashable}
	\item \url{https://medium.com/feed/startups-for-news}
\end{itemize}